



\magnification=1200
\vsize=22truecm
\hsize=15truecm
\voffset=0.5truecm
\hoffset=1truecm
\baselineskip=18truept
\lineskip=1pt
\lineskiplimit=0pt
\parskip=6truept


\font\titre=cmbx10 scaled\magstep1
\font\bigletter=cmr10 scaled\magstep2


\baselineskip=12pt

\line{\hfill McGill/92-30}
\line{\hfill hep-th/9205086}
\line{\hfill May 92}

\vskip 1in
\centerline {\titre Fractional Superspace Formulation of}
\centerline {\titre Generalized Super-Virasoro Algebras}
\vskip 0.75in
\centerline{{{\bigletter S}T\'EPHANE {\bigletter D}URAND}\footnote{$^{*}$}
{E-mail address: durand@hep.physics.mcgill.ca}}
\vskip 0.1in
\centerline{Department of Physics}
\centerline{McGill University}
\centerline{Ernest Rutherford Building}
\centerline{3600 University Street}
\centerline{Montr\'eal, PQ, Canada  H3A 2T8}
\vskip 0.75in

\centerline{\bf Abstract}
\vskip 0.1in

\noindent
We present a fractional superspace formulation of the centerless
parasuper-Viraso-ro and fractional
super-Virasoro algebras. These are two different generalizations
of the ordinary super-Virasoro algebra generated by the
infinitesimal diffeomorphisms of the superline. We work on the
fractional superline parametrized by $t$ and $\theta$, with
$t$ a real coordinate
and $\theta$ a paragrassmann variable
of order $M$ and canonical dimension $1/F$.
We further describe a more general structure
labelled by $M$ and $F$ with $M\geq F$.
The case $F=2$ corresponds to the parasuper-Virasoro
algebra of order $M$, while the case $F=M$ leads to the
fractional super-Virasoro algebra of order $F$.
The ordinary super-Virasoro algebra is
recovered at $F=M=2$.
The connection with $q$-oscillator algebras is discussed.

\vfill
\eject

\baselineskip=18truept

Symmetries play a fundamental role
in physics. Therefore, all new kinds of symmetry are worth studying. For
instance, supersymmetries could be a key ingredient of
unifying theories. Up to now, at least two generalizations of
supersymmetry are known: the para-supersymmetries$\,^{[1,2]}$ which
already possess interesting applications in quantum
mechanics$\,^{[1,3,4]}$, and
the fractional
supersymmetries$\,^{[5,6,7]}$ which appear, in particular, in the
context of the chiral Potts model$\,^{[8]}$. In this letter,
we shall present
a new algebraic stucture that allows the unification of the concepts
of para-superalgebra and fractional superalgebra, and that furthermore
describes new types of symmetries.
More precisely, we present a novel generalization of the centerless
super-Virasoro algebra which we will call a generalized super-Virasoro
algebra. This construction has the following features:

\noindent
$\bullet$ It is parametrized by 2 positive integers $M$ and $F$ with
$M\geq F$.

\parskip=0truept
\noindent
$\bullet$ It possesses a $Z_F$ grading.

\noindent
$\bullet$ It describes operators of fractional spin $s=1+{1\over F}$.

\noindent
$\bullet$ It is realized within the framework of a fractional
superspace formalism using paragrassmann variables of
order $M$ and canonical dimension $1/F$.

\noindent
$\bullet$ It contains as particular cases, the parasuper-Virasoro
algebras of order $M$ (for $F=2$) and the fractional super-Virasoro
algebras of order $F$ (for $M=F$). The ordinary super-Virasoro algebra
is recovered at $F=M=2$.

\parskip=6truept
\noindent
Thus, it also provides a new unified
realization of the ``old" fractional and para super-Virasoro
algebras. (In Ref.~$[2]$ the latter was realized in terms of the Green
representation$\,^{[9]}$ while in Refs.~$[5,6,7]$ the former was strictly
realized
in matrix form).

Let us first recall some standard results. The well-known centerless
super-Virasoro algebra is given by
$$\eqalignno{&[L_m,L_n] = (m-n)L_{m+n}\; &(1a)\cr
             &[L_n,G_r]=(\textstyle{1\over 2}n-r)G_{n+r}\; &(1b)\cr
             &\{G_r,G_s\} = 2L_{r+s}\; &(1c)\cr}$$
with $m,n\in Z$ and $r,s\in Z+{1\over 2}$. When centrally extended, it
yields the Neveu-Schwarz
algebra which plays a
central role in superstring theories. The algebraic
relations $(1)$, referred to as
the two-dimensional superconformal algebra,
can also be thought of as realized by
the generators of the infinitesimal diffeomorphisms
$(L_n)$ and superdiffeomorphisms $(G_r)$ of the superline:
$$\eqalignno{&L_n = t^{1-n}{\partial}_t - {\textstyle{1\over2}}(n-1)t^{-n}
\theta {\partial}_{\theta},\qquad n\in Z &(2a)\cr
             &G_r = t^{{1\over 2}-r}({\partial}_{\theta}+ \theta\partial_t),
 \qquad\qquad\qquad\;\> r\in Z+\textstyle{1\over2} &(2b)\cr}$$
where $t$ is the real line parameter
$({\partial}_t\equiv\partial/\partial t)$ and $\theta$ a
paragrassmann variable satifying
$({\partial}_{\theta}\equiv\partial/\partial\theta)$:
$$\theta^2=0, \qquad \{{\partial}_{\theta},\theta\}=1. \eqno(3)$$
The generators $L_n$ and $G_r$ respectively have spin $2$ and $3/2$.
Generally speaking,
one says that $\phi_n$ has conformal spin $s$
if
$$[L_m,\phi_n]=\big((s-1)m-n\big)\phi_{m+n}.  \eqno(4)$$
Note that $L_1$, $L_0$, $L_{-1}$, $G_{1/2}$ and $G_{-1/2}$ span an
$OSp(1,1)$ superalgebra and that $L_1$ and $G_{1/2}$ alone also
form a
superalgebra. We will come back to these subalgebras in the
context of quantum mechanics. Note also that the
algebra $(1)$ might have been presented with $(1c)$
replaced by
$$G_r^2=L_{2r} \eqno(5)$$
since $(1c)$ can be
reconstructed$\,^a$
from $(5)$ using $(1a,b)$.
In the following, the commutation relation $(1a)$ will stay unchanged, $(1b)$
will
keep the same form but $(1c)$ will be completely modified, being
replaced by a multilinear product.

We shall use the fractional superspace
formalism introduced in Ref.~$[10]$.
Very recently, a detailed analysis of this
formalism with a generalization to many paragrassmann
variables appeared in Ref.~[11].
We shall work on a space parametrized by a real coordinate $t$, a paragrassmann
variable $\theta$ of order $M$, and a $prime$ root of unity
$q$ $\in C$, satisfying
$$\theta^M=0, \qquad q^M=1 \qquad (q^n\not=1 ~{\rm for}~ n<M) \qquad
M=1,2,...\; . \eqno(6)$$
By a $prime$ root, we mean a root satisfying the condition in parenthesis,
$i.e.$, for a given order, we keep only roots which differ
from those of all lower orders ($e.g.$, $q\not=\pm 1$ for $M=4$). We also need
another parameter $F$ (called the fractional order) defined by
$$dim(\theta^F)=dim(t), \qquad F=1,2,...\; . \eqno(7)$$
Later, we will choose $M\geq F$.
The
derivative ${\partial}_{\theta}\equiv\partial/\partial\theta$,
which also satisfies ${\partial}_{\theta}^M=0$,
is introduced in the following way
$${\partial}_{\theta}\cdot\theta=1+q\theta{\partial}_{\theta}.\eqno(8)$$
We recover the ordinary grassmann case $(3)$ for
$M=2$ and the bosonic one for $M\rightarrow\infty$. The definition
$(8)$ implies
$${\partial}_{\theta}\cdot\theta^n = {1-q^n\over 1-q}\theta^{n-1}+q^n
\theta^n{\partial}_{\theta} \eqno(9)$$
with
$${1-q^n\over 1-q}=1+q+q^2+...+q^{n-1}
\qquad (=0~{\rm for}~n=M). \eqno(10)$$
A matrix realization of $\theta$ and $\partial_{\theta}$
is given in Ref.~$[11]$. Note that equation $(8)$,
which can be rewritten in the form
$$[{\partial}_{\theta},\theta]_q\equiv{\partial}_{\theta}\theta-
q \theta{\partial}_{\theta}=1, \eqno(11)$$
corresponds to one of the defining relations of the $q$-oscillator
algebra$\,^{[12]}$. We will return to this connection later.

We now introduce the generalized generators
$L_n^{(F;M)}$ and $G_r^{(F;M)}$ of the infinitesimal
diffeomorphisms of the fractional superline. They are of
fractional order $F$ and paragrassmann order $M$ with $M\geq F$:
$$L_n^{(F;M)}=t^{1-n}{\partial}_t
-\textstyle{1\over F}(n-1)t^{-n}B^{(M)},\qquad\qquad n\in Z\eqno(12a)$$
$$G_r^{(F;M)}=t^{{1\over F}-r}({\partial}_{\theta}+e \theta^{F-1}\partial_t)
-e(r-\textstyle{1\over F})t^{{1\over F}-r-1}\theta^{F-1}B^{(M)},\quad
r\in Z+\textstyle{1\over F}\eqno(12b)$$
with
$$B^{(M)}=\sum_{i=0}^{M-1}c_i\theta^i{\partial_\theta}^i
=c_0+\sum_{i=1}^{M-1}c_i\theta^i{\partial_\theta}^i \eqno(13)$$
and where
$$c_0:{\rm arbitrary}, \qquad c_i={(1-q)^i\over 1-q^i},
\qquad i=1,2,...\, ,M-1. \eqno(14)$$
Now, $\theta$
and $\partial_{\theta}$ satisfy relations (6,7,8). The
constant $e$ is not fixed for the moment.
The generators $(2)$ of the supersymmetric case are recovered
for $F=M=2$ and $c_0=0$. The operators $(12)$ satisfy the
following generalized super-Virasoro algebras:
$$\eqalignno{&[L_m,L_n] = (m-n)L_{m+n} &(15a)\cr
             &[L_n,G_r]=(\textstyle{1\over F}n-r)G_{n+r} &(15b)\cr}$$
$$G^{M}_r = \sum_{i=1}^{[\![M/F]\!]}
(-)^{i+1} a_i \, G_r^{M-Fi}L^i_{Fr} \eqno(15c)$$
with $L_n\equiv L_n^{(F;M)}$, $G_r\equiv G_r^{(F;M)}$, and where
$[\![M/F]\!]$ is the integer part of $M/F$. The coefficients
$a_i$ are given by
$$a_i\equiv a_i^{(F;M)}=
{M\over i}\left[{e\over (1-q)^{F-1}}\right]^i
b_i^{(F;M)}\eqno(16a)$$
where
$$b_i^{(F;M)}={(M-iF+i-1)!\over (i-1)!\,(M-i F)!} \ .\eqno(16b)$$
[The factor $(e/(1-q)^{F-1})^i$ in the coefficients $a_i$ may be
absorbed in the definition of the $L_n$.]
{}From $(4)$ and $(15b)$, we see that $G_r^{(F;M)}$ has fractional
spin $1+1/F$. Note that $a_1^{(F;M)}=eM/(1-q)^{F-1}$.
{}From now on, we set $e=(1-q)^{F-1}/M$ so that $a_1^{(F;M)}=1$.

The operator $B^{(M)}$ given in $(13)$ is such that, for
any order $M$, it satisfies:
$$[B^{(M)},\theta]=\theta, \qquad [B^{(M)},\partial_{\theta}]
=-\partial_{\theta}. \eqno(17)$$
This makes contact with quantum algebras
since eqs. $(11)$ and $(17)$ are actually a particular case
(for $i=j=1$ and $b_1=\partial_{\theta}$, $b_1^\dagger=\theta$) of
the defining relations of multi $q$-oscillator algebra$\,^{[12]}$:
$$[b_i,b_i^{\dagger}]_q=1, \qquad [b_i,b_j^{\dagger}]=0 \quad (i\not=j),
\qquad[N,b_i]=-b_i,
\qquad [N,b_i^{\dagger}]=b_i^{\dagger}. \eqno(18)$$
The formula $(17)$ allows one to compute
exactly the commutation relation $(15b)$. However, $(15c)$ has been verified
only for
particular $F$ and $M$.

Let us point out some limiting cases of formulas $(15c)$.
First, for $F=M$ we recover the fractional super-Virasoro
algebras of order $F$:
$$G_r^F=L_{Fr}\;.   \eqno(19)$$
The trivial
case $F=1$ shows that $G_r^{(1;1)}=eL_r^{(1;1)}$. At $F=2$, we
recover the supersymmetric case $(5)$ which is equivalent
to $(1c)$ as already mentionned. Actually, this kind of equivalence
exists for all orders $F$. Indeed, from $(15a,b)$ and $(19)$, one may
derive$\,^b$ the general formula
$$\{G_{r_1},G_{r_2},\,...\,,G_{r_F}\}=F!\,\,L_{r_1+r_2+...+r_F}\eqno(20)$$
where we have used the following definition of the multilinear
symmetric product
$$\{X_1,X_2,\,...\,,X_M\}\equiv X_1X_2...X_M
+\; all \;\, permutations \;\, of \;\, the \;\, X_i.\eqno(21)$$
(For instance, for $M=3$ the right hand side of $(21)$ contains
6 terms.) This symmetric product is a kind of generalization of
the anti-commutator and $(1c)$ is a particular case of $(20)$.
Eqs.~$(15a,b)$ and $(20)$ define
the centerless fractional super-Virasoro algebras$\,^{[5,6,7]}$
which have a $Z_F$ grading. (In Ref.~$[7]$, these algebras are
presented in an interesting different form.) Finally, these structures
possess a subalgebra generated by
$Q\equiv G_{1/F}=({\partial}_{\theta}+e\theta^{F-1}\partial_t)$
and $H\equiv L_1=\partial_t$:
$$Q^F=H, \qquad [H,Q]=0. \eqno(22)$$
Thus $Q$ appears as a fractional covariant derivative. For $F>2$,
this is the only subalgebra (actually, we may add $L_0$). This subalgebra has
arisen in the
context of the chiral Potts model$\,^{[8]}$.

The second limiting case consists of the parasuper-Virasoro algebras
of order $M$ which are recovered at $F=2$. Here, it is more difficult
to present these algebras in a simpler form than in $(15c)$; therefore
we shall make them explicit for a few cases
(we have set $a_i^{(2;M)}\equiv a_i^{(M)}$):
$$\eqalignno{&M=2: \qquad\quad G_r^2=      L_{2r}
                      \cr
             &M=3: \qquad\quad G_r^3=G_r   L_{2r}
\cr
             &M=4: \qquad\quad G_r^4=G_r^2 L_{2r}-a_2^{(4)}     L_{2r}^2
	                    &(23)\cr
             &M=5: \qquad\quad G_r^5=G_r^3 L_{2r}-a_2^{(5)}G_r  L_{2r}^2
	             \cr
             &M=6: \qquad\quad G_r^6=G_r^4
L_{2r}-a_2^{(6)}G_r^2L_{2r}^2+a_3^{(6)}L_{2r}^3  \cr
             &\;\; etc. \cr}$$
As before, we can reconstruct$\,^c$ the multilinear product from
$(15a,b)$ and $(23)$. For instance, we find
$$\eqalignno{&M=3: \qquad \{G_{r},G_{s},G_{t}\}=G_{r}L_{s+t}+\; perm.
     &(24a)\cr
             &M=4: \qquad \{G_{r},G_{s},G_{t},G_{u}\}=G_{r}G_{s}L_{t+u}+
a_2^{(4)}L_{r+s}L_{t+u}+\; perm.\qquad &(24b)\cr}$$
Now ``$perm.$" signifies that one should add the terms
which are obtained from those already appearing
in the right-hand side of $(24)$ by performing all
permutations of $(r,s,t,...)$.
The structure relations $(15a)$, $(15b)$ for $F=2$, and $(24)$ define the
centerless
parasuper-Virasoro algebras which have a $Z_2$ grading.
(The coefficients $a_i^{(M)}$ are not the same as in Ref.~[2].)
The smallest subalgebra is generated as before by $Q\equiv
G_{1/2}=({\partial}_{\theta}+e\theta\partial_t)$ and
$H\equiv L_1=\partial_t$. From $(15b)$ and $(23)$
we find $[H,Q]=0$ and $Q^2=H$ for $M=2$, $Q^3\cong QH$ for $M=3$,
$Q^4\cong Q^2H-H^2$ for $M=4$,
$etc$, which are the defining relations of the parasupersymmetric
quantum mechanics of order $M$ $^{[4]}$, where $Q$ is the
parasupercharge and $H$ the hamiltonian. Actually, there exists
a larger subalgebra given by $L_1$, $L_0$, $L_{-1}$, $G_{1/2}$
and $G_{-1/2}$ (this is not true for $F>2$) which spans the paragrassmann
generalization of the $OSp(1,1)$ superalgebra. In this regard, notice that
the parasuperconformal quantum mechanics of order $M=3$ has been
constructed in Ref. $[3]$.

All the other cases $(M\not=F\not=2)$ are new symmetry algebras.
For instance, the intermediate case $F=3$ and $M=6$ reduces
for $r=1/3$ to
$$Q^6=a_1 Q^3H-a_2 H^2 \eqno(25)$$
with the $a_i$ given by $(16a)$.
It is also possible to generate a multi-index version of $(15c)$
for all of these novel cases.

Before concluding, let us write an interesting formula.
For a given $q$ of order $M$, one has
$$\sum_{(n_1,\ldots,n_i)\in S_i} \ \prod_{j=1}^i \;(1-q^{n_j})=
\left(\matrix{M\cr i\cr}\right)
\equiv{M!\over i!\,(M-i)!}\eqno(26)$$
for $S_i$ the set of combinations of $i$ numbers out of the set
$\{1,2,\,...\,,M-1\}$.
As particular cases, one finds for $i=1$ and $i=M-1$:
$$\sum_{j=1}^{M-1}(1-q^j)=\prod_{j=1}^{M-1}(1-q^j)=M. \eqno(27)$$

The next natural step is to compute the central
extension of these generalized super-Virasoro algebras. Maybe, one would find
central terms with a dependence
on $F$, $M$ and $D$, leading to new critical
dimensions $D$ for $M\not=2$.

\vskip 0.5cm
\centerline{\bf Acknowledgments}
\vskip 0.2cm

I am pleased to thank Luc Vinet and Vyacheslav Spiridonov for
their interest and useful discussions.
This work is supported in part through funds provided by the Natural Sciences
and
Engineering Research Council (NSERC) of Canada and the Fonds FCAR of the
Qu\'ebec Ministry of Education.

\vskip 0.7cm
\centerline{\bf Notes}
\vskip 0.2cm

\par
\frenchspacing

\item{$^{(a)}$}
More explicitly, (1c) is reconstructed from $[L_{r-s},G^2_s] =
[L_{r-s},L_{2s}]$. Strictly
speaking, we should also add
$\{ G_s,G_{3s}\} = 2L_{4s}$ to $(5)$. However, this equation
follows if we
assume that the (anti--)commutation relations have a regular
structure.

\item{$^{(b)}$}
For instance, for $F=3$ we must
successively commute each side of the equation $(19)$
with $L_{s-r}$ and $L_{t-r}$. Strictly
speaking, we should here add
$\{ G_r,G_{4r},G_{4r}\}=6L_{9r}$ to $(19)$.
This equation follows if we assume a regular structure.

\item{$^{(c)}$}
Now, for $M=3$ for instance, we must perform the same trick
on $(23)$ as the one mentioned in the preceding note but
adding the relation
$(24a)$ with $s=3r$ and $t=3r$.

\par
\vfill\eject

\centerline{\bf References}
\vskip 0.2cm
\par
\frenchspacing

\item{[1]}
V.A. Rubakov and V.P. Spiridonov, {\it Mod. Phys. Lett.} {\bf A3}, 1337 (1988).

\item{[2]}
S. Durand, R. Floreanini, M. Mayrand and L. Vinet, {\it Phys. Lett.}
{\bf B233}, 158 (1989).

\item{[3]}
S. Durand and L. Vinet, {\it Mod. Phys. Lett.} {\bf A4}, 2519 (1989);
S. Durand and L. Vinet, in {\it Field Theory and Particle Physics},
edited by O. J. P. Eboli, M. Gomes and A. Santoro
(World Scientific, Singapore, 1990) pp. 291-314.
For a review, see Ref. [6].

\item{[4]}
S. Durand, M. Mayrand, V.P. Spiridonov and L. Vinet, {\it Mod. Phys. Lett.}
{\bf A6}, 3163 (1991).

\item{[5]}
S. Durand, {\it Parasupersym\'etrie}, Ph.D. thesis, University of Montr\'eal
(1990).

\item{[6]}
S. Durand, R. Floreanini, M. Mayrand, V.P. Spiridonov and L. Vinet,
``Parasupersymmetries and Fractional Supersymmetries",
in {\it Quantum Field Theory, Quantum Mechanics and Quantum Optics},
Proceedings of the  XVIIIth International Colloquium on Group
Theoretical Methods in Physics, 4-9 June 1990, Moscow, USSR,
edited by V.V. Dodonov and V.I. Man'ko
(Nova Science Publishers, 1991) pp. 145-154.

\item{[7]}
T. Nakanishi, {\it Progr. Theor. Phys.} {\bf 82}, 207 (1989);
{\it Mod. Phys. Lett.} {\bf A3}, 1507 (1988).

\item{[8]}
D. Bernard and V. Pasquier, Saclay preprint SPhT/89-284 (1989); D. Bernard and
A. Leclair, Saclay preprint SPhT/90-082 (1990).

\item{[9]}
Y. Ohnuki and S. Kamefuchi, {\it Quantum Field Theory and Parastatistics},
(Springer-Verlag, New York, 1982).

\item{[10]}
C. Ahn, D. Bernard and A. Leclair, {\it Nucl. Phys.} {\bf B346}, 409 (1990).

\item{[11]}
A.T. Filippov, A.P. Isaev and A.B. Kurdikov, ``Paragrassmann Analysis and
Quantum Groups", JINR Dubna preprint (April 1992) hep-th/9204089.

\item{[12]}
See for instance: R. Floreanini and L. Vinet, {\it Lett. Math. Phys.}
{\bf 22}, 45 (1991); ``Representations of
Quantum Algebras and $q$-Special Functions" in {\it IIth International Wigner
Symposium}, edited by V. Dobrev and H. Dobner (Springer-Verlag, Berlin, 1992);
V.P. Spiridonov, private communication.

\par
\vfill
\end